\documentclass[12pt]{iopart}

\usepackage{graphicx}
\usepackage{iopams}
\usepackage{setstack}

\newcommand{\eqref}[1]{(\ref{#1})}

\begin{document}

\title[Ratchet effect on a relativistic particle]{Ratchet effect  
on a relativistic particle driven by external forces}

\author{Niurka R.\ Quintero$^1$, Renato Alvarez-Nodarse$^2$ and
Jos\'e A.\ Cuesta$^3$}

\address{$^1$ Departamento de F\'\i sica Aplicada I, E.\ S.\ P.,
Universidad de Sevilla, C/ Virgen de \'Africa 7, E-41011, Sevilla, Spain}
\address{$^2$ Departamento de An\'alisis Matem\'atico, Universidad de Sevilla,
apdo.~1160, E-41080, Sevilla, Spain}
\address{$^3$ Grupo Interdisciplinar de Sistemas Complejos (GISC), Departamento
de Matem\'aticas, Universidad Carlos III de Madrid, avda.~de la Universidad 30,
E-28911 Legan\'es, Madrid, Spain}

\eads{\mailto{niurka@us.es}, \mailto{ran@us.es}, \mailto{cuesta@math.uc3m.es}}

\begin{abstract}
We study the ratchet effect of a damped relativistic particle driven by both
asymmetric temporal bi-harmonic and time-periodic piecewise constant forces.
This system can be formally solved for any external force, providing the
ratchet velocity as a non-linear functional of the driving force. This allows
us to explicitly illustrate the functional Taylor expansion formalism
recently proposed for this kind of systems. The Taylor expansion reveals
particularly useful to obtain the shape of the current when the force is
periodic, piecewise constant. We also illustrate the somewhat
counterintuitive effect that introducing damping may induce a ratchet
effect. When the force is symmetric under time-reversal and the system is
undamped, under symmetry principles no ratchet effect is possible. In this
situation increasing damping generates a ratchet current which, upon
increasing the damping coefficient eventually reaches a maximum and 
decreases toward zero. We argue that this effect is not specific of this
example and should appear in any ratchet system with tunable damping
driven by a time-reversible external force.
\end{abstract}

\submitto{\JPA}

\maketitle

\section{Introduction}

The ratchet effect is identified with the motion of particles or solitons 
induced by zero-average periodic forces \cite{reimann:2002a,salerno:2002a},
sometimes in the presence of thermal fluctuations. The effect arises as a
subtle interplay between nonlinearities in the system and broken 
symmetries. Ratchets appear in many fields of physics, where net motion is
generated either by an asymmetric, periodic, spatial potential
\cite{magnasco:1993,falo:2002,reimann:2002b,astumian:2002,linke:2002,villegas:2003,%
beck:2005,hanggi:2009}, or by an asymmetric temporal forcing
\cite{hanggi:2009,schneider:1966,ajdari:1994,engel:2003,morales-molina:2003,%
schiavoni:2003,ustinov:2004,cole:2006,ooi:2007}.  
In both cases the ratchet effect can be regarded as an application of
Curie's symmetry principle, which states that a symmetry transformation
of the cause (forces) is also a symmetry transformation of the effect
(ratchet velocity) \cite{curie:1894,ismael:1997}.   

Most studies of ratchets driven by temporal forces employ a bi-harmonic
forcing
\begin{eqnarray}
f(t)=\epsilon_{1} \cos(q \omega t +\phi_{1})+
\epsilon_{2} \cos(p \omega t +\phi_{2}), 
\label{eq-f}
\end{eqnarray}
where $p$ and $q$ are positive integers which, without loss of generality,
can be taken co-prime (otherwise common factors can be absorbed in
the frequency $\omega$) and $\epsilon_1$, $\epsilon_2$ are small non-zero
parameters. If both $p$ and $q$ are odd, the force \eqref{eq-f}
exhibits the \emph{shift symmetry} $(\mathcal{S}f)(t)=f(t+T/2)=-f(t)$,
where $T=2\pi/\omega$.
In systems invariant under time translations this implies that both,
$f(t)$ and $-f(t)$ generate the same ratchet current (or velocity) defined 
as \cite{flach:2000,salerno:2002}
\begin{eqnarray}
v = \lim_{t \to +\infty} \frac{1}{t} \,
\int_{0}^{t}  \dot x(\tau) \, d\tau = \lim_{t \to +\infty} \frac{x(t)}{t},
\label{eqav}
\end{eqnarray}
where $x(t)$ is the position of the particle, soliton, or localized structure.
If reversing the force changes the sign of the current, this current must be zero.
So shift-symmetric bi-harmonic forces cannot induce a ratchet effect.
In contrast, if $p$ and $q$ have different parity, shift symmetry
is broken by $f(t)$ so the force can induce a nonzero net current
\cite{quintero:2010}.

Many attempts have been made to determine quantitatively the dependence
of the ratchet velocity, $v$, on the parameters of the bi-harmonic force
\eqref{eq-f} \cite{schneider:1966,skov:1964,denisov:2002}.
Invariably, the analysis performed in these works rests on the so-called
\emph{method of moments,} where it is assumed that the average ratchet
velocity can be expanded as a series of the odd moments of $f(t)$, i.e. 
$\sum_{k=1}^{\infty} \langle [f(t)]^{2 k +1} \rangle$ with 
$\langle h(t) \rangle= \int_{0}^{T} dt \, h(t)$. This
method seemed to work for some systems but not for others without a
clear reason and with no known criterion to tell ones from the others.
We have recently shown that the moment method relies on an assumption
that almost never holds, and have provided an alternative procedure that
yields the correct result regardless of the system \cite{quintero:2010}.

The aim of this paper is to provide explicit examples which illustrates
this otherwise abstract method ---the functional expansion of $v$ in
terms of $f$--- using a working example for which an analytic
solution can be found. The system represents the motion of a damped,
relativistic particle under the effect of two different forces: a bi-harmonic
force like \eqref{eq-f}, and a time-periodic piecewise constant force like
\begin{equation}
f(t)=
\cases{
\epsilon_1 & \text{if $0<t<T_l$,} \\
0 & \text{if $T_l< t< T-T_{l}$,} \\
-\epsilon_1 & \text{if $T-T_l<t<T$.}
}
\label{sq-wv}
\end{equation}
To this purpose we introduce the model as well as its
analytic solution in section~\ref{sec2}. In section~\ref{sec:damping}
we discuss the phenomenon of damping-induced ratchets. The formalism
developed in \cite{quintero:2010} is fully illustrated for this
problem in section~\ref{sec3}. 
For these two specific driving forces it is also shown that 
the method of moments is valid only when the dynamics of the relativistic
particle is overdamped, and fails otherwise.
Conclusions are summarized in section~\ref{sec:discussion}.

\section{Motion of a relativistic particle driven by a bi-harmonic force}
\label{sec2}

The equation of motion of a relativistic particle with mass $M>0$, whose
position and velocity at time $t$ are denoted $x(t)$ and $u(t)$,
respectively, is
\numparts
\label{equx}
\begin{eqnarray}
&\frac{dx}{dt} = u(t), &\qquad x(0)=x_{0}, \\
M &\frac{du}{dt} = - f(t) (1-u^2)^{3/2}-\gamma u(1-u^2), &
\qquad u(0)=u_0,\label{equxa}
\end{eqnarray}
\endnumparts
where $x_{0}$ and $u_{0}$ are the initial conditions, $\gamma>0$ represents 
the damping coefficient and $f(t)$ is a $T$-periodic driving force.     
Notice that if the force $f(t)$ satisfied $(\mathcal{S}f)(t)=f(t+T/2)=-f(t)$,
then \eref{equxa} would be invariant under a combination of shift symmetry
($\mathcal{S}\,:\, t\mapsto t+T/2$) and the sign change $x\mapsto -x$.

Changing the variable $u(t)$ by the momentum
\begin{equation}
\label{def-P}
P(t)=\frac{M u(t)}{\sqrt{1-u^2(t)}}
\end{equation}
transforms \eref{equxa} into the linear equation
\begin{equation}
\label{eq-P}
\frac{dP}{dt}=-\beta P-f(t),\quad P(0)=P_0=\frac{M u_0}{\sqrt{1-u^2_0}},
\end{equation}
where $\beta=\gamma/M$. Equation~\eqref{eq-P} is easily solved to give
\begin{equation} \label{solu-P}
P(t) = P_0 \rme^{-\beta t} - \int_{0}^{t} \, \rmd z f(z) \rme^{-\beta (t-z)}.
\end{equation}
From \eqref{def-P} one obtains
\begin{eqnarray} 
\label{eq-u1}
u(t) &= & \sum_{k=0}^{\infty} \left(-\frac{1}{2}\right)^{k}\frac{(2k-1)!!}{k!} 
\left(\frac{P(t)}{M}\right)^{2 k +1}.
\end{eqnarray}

Let us now focus our attention on the 
$T$-periodic driving force $f(t)$ given by (\ref{eq-f}) with $p=2$ and $q=1$ 
(the most common choice of parameters
\cite{olsen:1983,salerno:2002,morales-molina:2003,morales-molina:2006}).
Substituting \eref{eq-f} into \eref{solu-P} leads to
\begin{eqnarray*}
P(t) = \widetilde{P}_{0} \rme^{-\beta t}-\tilde\epsilon_1
\cos(\omega t+\phi_{1}-\chi_{1}) 
 -\tilde\epsilon_2 \cos(2 \omega t+\phi_{2}-\chi_{2}),
\end{eqnarray*}
with 
\begin{eqnarray*}
\widetilde{P}_{0} &=P_0+\tilde\epsilon_{1}\cos(\phi_{1}-\chi_{1}) + 
\tilde\epsilon_{2}\cos(\phi_{2}-\chi_{2}), & \\
\tilde\epsilon_1&=\epsilon_{1}(\beta^2+\omega^2)^{-1/2}, 
&\chi_{1}=\tan^{-1}\left(\omega/\beta\right), \\
\tilde\epsilon_2&=\epsilon_{2}(\beta^2+4 \omega^2)^{-1/2}, &
\chi_{2}=\tan^{-1}\left(2 \omega/\beta\right). 
\end{eqnarray*}

As $t\to\infty$ the momentum $P(t)$ behaves, for any $\beta>0$, as
\begin{equation*}
P(t)\sim -\tilde\epsilon_1
\cos(\omega t+\phi_{1}-\chi_{1}) 
 -\tilde\epsilon_2 \cos(2 \omega t+\phi_{2}-\chi_{2}),
\end{equation*}
thus the term $P(t)^{2k+1}$ in \eqref{eq-u1} is $O(\epsilon_1^r
\epsilon_2^s)$ with $r+s=2k+1$. Since the time average of $P(t)$ is zero,
the leading term of \eqref{eqav} in powers of $\epsilon_1$ and
$\epsilon_2$ will be
\begin{eqnarray*}
\fl
-\frac{1}{2M^3}\lim_{t\to\infty}\frac1t\int_0^tP(\tau)^3\,\rmd \tau &=\frac{3}{2M^3T}
\tilde\epsilon_1^2\tilde\epsilon_2
\int_0^T\cos(\omega \tau+\phi_{1}-\chi_{1})^2\cos(2 \omega \tau+\phi_{2}-\chi_{2})\,
\rmd \tau \\
 &=\frac{3}{8M^3}\tilde\epsilon_1^2\tilde\epsilon_2
\cos(2\phi_1-\phi_2+\chi_2-2\chi_1).
\end{eqnarray*}
Therefore, the rachet velocity \eref{eqav}, for small 
amplitudes $\epsilon_{1}$ and $\epsilon_{2}$, is given by
\begin{equation}
\label{vmedia-1}
v = B \epsilon_1^2\epsilon_2  \cos(2\phi_1-\phi_2+\theta_{0}),   
\label{eq:vrelat}
\end{equation}
with
\begin{equation}
\fl
B=\frac{3}{8M^3(\beta^2+\omega^2) \sqrt{\beta^2+4\omega^2}},
\qquad \theta_{0}=\chi_2-2\chi_1=-\tan^{-1}\left(\frac{2\omega^3}{\beta
(\beta^2+3\omega^2)}\right),
\label{coeff}
\end{equation}
in agreement with the result reported in \cite{quintero:2010}.

Notice that in the undamped limit $\gamma\to 0$ (equivalently $\beta\to 0$)
the parameters \eqref{coeff}
become
\begin{equation*}
B=\frac{3}{16M^3\omega^3}, \qquad \theta_0=-\frac{\pi}{2},
\end{equation*}
whereas in the overdamped limit $M\to 0$ (and therefore $\beta\to\infty$
with $M\beta=\gamma$) we get
\begin{equation*}
B=\frac{3}{8\gamma^3}, \qquad \theta_0=0,
\end{equation*}
both limits agree with the predictions of \cite{quintero:2010}.

\section{Ratchet induced by damping}
\label{sec:damping}

\begin{figure}
\centerline{\includegraphics[width=10cm]{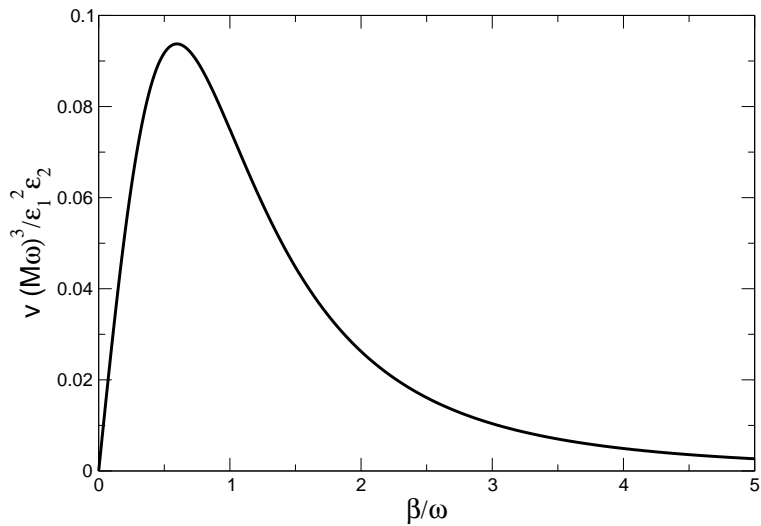}}
\caption[]{\label{fig:damping}
Plot of the current velocity $v$, in units of $\epsilon_1^2\epsilon_2
/(M\omega)^3$, vs.\ the damping coefficient $\beta$, in units of the
frequency, $\omega$, induced by a biharmonic force like \eqref{eq-f}
with $\phi_1=\phi_2=0$. Notice that this force is time reversible,
i.e., $f(-t)=f(t)$.}
\end{figure}

The depence of $v$ on parameters of the system like the damping coefficient
(through $\beta=\gamma/M$) shown in \eqref{eq:vrelat} and \eqref{coeff}
reveals an interesting effect. If we take $\phi_{1}=\phi_{2}=0$ in $f(t)$
and do some algebra, the ratchet velocity (for small amplitudes of
the force) turns out to be
\begin{equation}
v=\frac{\epsilon_1^2\epsilon_2}{\omega^3M^3}V(\beta/\omega), \qquad
V(x)=\frac{3x(x^2+3)}{8(x^2+1)^2(x^2+4)}.
\end{equation}
Function $V(x)$ is depicted in Figure~\ref{fig:damping}. The most remarkable
observation is that the current \emph{increases} up to a maximum with
increasing damping before it begins to show the expected decay. Intuition
dictates that the current should decrease with damping because friction
opposes movement, so the fast increase it reveals for small damping is
counterintuitive.

The cause of this effect is the interplay between the breaking of the
time-reversal symmetry $\mathcal{R}\,:\, t\mapsto -t$ that generates
the ratchet current, and the damping that hinders it
\cite{morales-molina:2006}. In the limit
$\beta\to 0$ the system \eqref{equx} is invariant under $\mathcal{R}$
and a sign change of $u$, because for $\phi_1=\phi_2=0$ the force
\eqref{eq-f} satisfies $f(-t)=f(t)$. Accordingly $v=0$ in this limit.
But introducing damping breaks the symmetry of the equation and induces
a net movement of the particle. For small damping, the higher the
damping coefficient $\beta$ the larger $v$. If we keep on increasing
$\beta$ eventually the friction it introduces in the movement of the
particle causes the decay of $v$ as $\beta^{-3}$.

This argument makes it clear that in any ratchet system with a tunable
damping and undergoing the action of a time-reversible bi-harmonic force,
the ratchet effect can be generated upon increasing damping above zero.
   
\section{Ratchet velocity as a functional of the force}
\label{sec3}

The starting point to obtain formula \eqref{vmedia-1} for a ratchet system
is to realize that $v$ is a functional of $f(t)$ and that, under certain
regularity assumptions, one such functional can be expanded as a functional
Taylor series \cite{curtain:1977,binney:1992,hansen:2006} as
\begin{equation}\label{eq-ov2}
v[f]= \sum_{n{\rm\ odd}} \int_0^T \frac{\rmd t_1}{T}\cdots
\int_0^T \frac{\rmd t_{n}}{T} 
c_{n}(t_1,\dots,t_{n})f(t_1)\cdots f(t_{n}),
\end{equation} 
where the kernels $c_{n}(t_1,\dots,t_{n})$ are proportional to the $n$th
functional derivatives of the functional $v[f]$. These kernels can be taken
$T$-periodic in each variable and totally symmetric under any exchange of
variables. Only odd terms appear in this expansion as a cosequence of
the symmetry $v[-f]=-v[f]$ that these systems have.

That $v$ is indeed a functional of $f(t)$ in this example is obvious from
equations~\eqref{eqav}--\eqref{solu-P}. The aim of this section is
to determine explicitly the expansion \eqref{eq-ov2} for this exactly
solvable example.

Let us start off by rewriting the integral in \eref{solu-P} as
\begin{eqnarray}  \label{eq-I} 
 \int_{0}^{t} \rmd z f(z)  \rme^{-\beta (t-z)} = 
 I_{1}(t) + I_{2}(t),  \\ \nonumber 
  I_{1}(t) = \sum_{k=1}^{n(t)} \int_{0}^{T} \rmd z f(z)  \rme^{-\beta (t-z-(k-1)T)}, \\ 
\nonumber 
  I_{2}(t) = \int_{0}^{\alpha(t)} \rmd z f(z)  \rme^{-\beta (\alpha(t)-z)}, 
\end{eqnarray}
where $\alpha(t)=t- n(t)T$ and $n(t)=[t/T]$ ($[X]$ denoting the integer part
of $X$). Notice that $\alpha(t+T)=\alpha(t)$. Now, since 
\begin{equation} 
S(t)\equiv\sum_{k=1}^{n(t)} \rme^{\beta (k-1)T} =
\frac{\rme^{\beta n T}-1}{\rme^{\beta T}-1}, 
\end{equation}
then $I_{1}(t) =  \rme^{-\beta t}\,C\,S(t)$, with
\begin{equation}
C=\int_{0}^{T} \rmd z f(z) [\rme^{\beta z}-1].
\end{equation}
Using this form in \eqref{solu-P} we can write
\begin{equation} \label{solu-P1}
P(t) = A \rme^{-\beta t} + \widetilde{P}(t), 
\end{equation}
where $A = P(0) + C(\rme^{\beta T}-1)^{-1}$
and $\widetilde{P}(t)$ is the $T$-periodic function
\begin{eqnarray} \nonumber
\widetilde{P}(t) = - \frac{1}{\rme^{\beta T}-1} \int_{0}^{T} \, \rmd y
f(y) \rme^{-\beta \alpha(t)} [\rme^{\beta y}-1] 
-\int_{0}^{\alpha(t)} \, \rmd y f(y) \rme^{-\beta (\alpha(t)-y)}.
\label{solu-Pt}
\end{eqnarray}
It is thus enough to obtain $\widetilde{P}(t)$ in the interval $0 \le t <T$,
where it adopts the compact form  
\begin{equation} \label{pta}
\widetilde{P}(t) =  - \int_{0}^{T} \, \rmd y f(y) \rme^{-\beta (t-y)} \chi(y,t),
\end{equation}
defining 
\begin{equation} \label{chi0}
\chi(y,t) = \frac{1-\rme^{-\beta y}}{\rme^{\beta T}-1} + \Theta(t-y)
\end{equation}
(as it is customary, $\Theta(x)$ denotes the Heaviside function, which is
$1$ if $x>0$ and $0$ otherwise).

Equations~\eqref{pta}--\eqref{chi0} have a well defined $\beta\to 0^+$ limit,
namely
\begin{equation} \label{chi1}
\fl
\widetilde{P}(t) =  - \int_{0}^{T} \, \rmd y f(y) \chi_1(y,t),\qquad 
\chi_1(y,t) = \frac{y}{T} + \Theta(t-y).
\end{equation} 
On the other hand, for zero-average forces $f(t)$ the kernel $\chi(t,z)$
can be further simplified to
\begin{equation} \label{chi2}
\fl
\widetilde{P}(t) =  - \int_{0}^{T} \, \rmd y f(y) \rme^{-\beta (t-y)} \chi_2(y,t),\qquad
\chi_2(y,t) = \frac{1}{\rme^{\beta T}-1} + \Theta(t-y).
\end{equation} 
Whatever the form, it should be periodically extended beyond the interval
$[0,T)$.

It is then clear that \eqref{eqav} and \eqref{eq-u1} boil down to
\begin{equation} \label{eqav1}
 v =  \sum_{k=0}^{\infty} \left(-\frac{1}{2}\right)^k\frac{(2k-1)!!}{k!
\,M^{2k+1}} \int_{0}^{T} \frac{\rmd \tau}{T} \widetilde{P}(\tau)^{2 k +1}.
\end{equation}
A direct comparison of this equation with the functional Taylor series
\eqref{eq-ov2} yields
\numparts
\begin{eqnarray}
&c_{2k}(t_1,\dots,t_{2k}) =  0 , \\
&c_{2k+1}(t_1,\dots,t_{2k+1})  = \left(-\frac{1}{2}\right)^k
\frac{(2k-1)!!}{k! \,M^{2k+1}}
T^{2k} a_{2k+1}(t_1,\dots,t_{2k+1}),
  \label{eq-ck}  
\end{eqnarray}\endnumparts
 where
\begin{equation}\label{eq-ak}
a_{m}(t_1,\dots,t_{m}) =
\int_{0}^{T} \rmd\tau \,\rme^{-\beta m (\tau-\bar{t})}
\prod_{k=1}^{m} \chi(t_{k},\tau),
\qquad \bar{t}=\frac{1}{m}\sum_{k=1}^{m} t_{k}.
\end{equation}  
As expected \cite{quintero:2010}, functions $a_{m}(t_1,\dots,t_{m})$ are,
by construction, $T$-periodic
in each variable and symmetric under any exchange of their arguments.

The integral in \eqref{eq-ak} can be performed integrating by parts and
taking into account that $\frac{d}{d\tau}\chi(y,\tau)=\delta(\tau-y)$
(a Dirac delta). The result is
\begin{equation}
\fl
a_{m}(t_1,\dots,t_{m})=\frac{e^{\beta m\bar t}}{\beta m}\left\{
\prod_{k=1}^m \chi(t_k,0)-\prod_{k=1}^m\chi(t_k+T,0)+
\sum_{j=1}^m e^{-\beta mt_j} \prod_{k=1,\ k\neq j}^m\chi(t_k,t_j)
\right\},
\end{equation}
where we have used the fact that $\chi(t_k,T)e^{-\beta T}=
\chi(t_k+T,0)$. As usual, empty products are assumed to be $1$ (the case
of the last term for $m=1$).

The limit $\beta \to 0$ of this expression is better obtained by
replacing $\chi(y,t)$ by $\chi_1(y,t)$ in \eqref{eq-ak} and integrating
by parts again. This results in
\begin{equation}\label{eq-ak1}
a_{m}(t_1,\dots,t_{m})=T\prod_{k=1}^m \chi_1(t_k,T)-
\sum_{j=1}^m t_j \prod_{k=1,\ k\neq j}^m\chi_1(t_k,t_j).
\end{equation} 

Finally, in the overdamped case ($M\to 0$, $\beta\to\infty$), instead
of \eref{eq-P} the evolution of $P$ is given by $P(t)=-(1/\beta) f(t)$,
so $v$ can be expressed simply as
\begin{equation}\label{eq-ov1}
v = - \sum_{k=0}^{\infty} \left(-\frac{1}{2}\right)^k \frac{(2k-1)!!}{k!
\, \gamma^{2k+1}} \frac{1}{T} \int_{0}^{T}\, \rmd t\, f(t)^{2k+1}.
\end{equation}  
From \eqref{eq-ov2} and \eqref{eq-ov1} it follows that
$c_{2k}(t_1,\dots,t_{2k})=0$ and
\begin{equation}\label{eq-ov3}
\fl
c_{2k+1}(t_1,\dots,t_{2k+1}) = -\left(-\frac{T^2}{2}\right)^k
\frac{(2k-1)!!}{k! \, \gamma^{2k+1}}\,
\delta(t_1-t_{2})\cdots  \delta(t_{2k}-t_{2k+1}). 
\end{equation}   

\subsection{Forcing with a time-periodic piecewise constant force}
\label{sec4}
The expansion \eqref{eq-ov2} with kernels \eqref{eq-ck} and \eqref{eq-ak}
turns out to be useful to analyze different types of forcing. For instance,
another standard choice in the literature (see \cite{reimann:2002a} and references therein), 
alongside with the bi-harmonic force \eqref{eq-f}, has been the time-periodic piecewise 
constant force defined in \eqref{sq-wv}.
This force is shift-symmetric only for $T_l=T/2$, so any other value $T_l<T/2$
breaks this symmetry and induces a ratchet current.

In order to ascertain the effect of this force in system \eqref{equxa}
for small amplitudes $\epsilon_1\ll 1$, we will compute the first nonzero
term in the expansion \eqref{eq-ov2}. To that purpose we need to evaluate
(c.f.~equation~\eqref{eq-ak})
\numparts
\begin{eqnarray}
\label{eq:amItau}
&K_m\equiv\langle a_m(t_1,\dots,t_m)f(t_1)\cdots f(t_m)\rangle =\int_0^T
\left[e^{-\beta\tau}I(\tau)\right]^m d\tau, \\
&I(\tau)\equiv\frac{1}{T}\int_0^Te^{\beta t}\chi_2(t,\tau)f(t)\,dt,
\label{eq:Itau}
\end{eqnarray}
\endnumparts
where the choice $\chi_2(t,\tau)$ instead of $\chi(t,\tau)$ is made
because $f(t)$ in \eqref{sq-wv} has zero average. According to \eqref{chi2}
$\chi_2(t,\tau)=(1-e^{-\beta T})^{-1}\bar\chi_2(t,\tau)$, where
\begin{equation}
\bar\chi_2(t,\tau)=
\cases{
1 & \text{if $t<\tau$,} \\
e^{-\beta T} & \text{if $t>\tau$.}
}
\end{equation}
Substitution into \eqref{eq:Itau} yields
\numparts
\begin{eqnarray}
I(\tau)=\frac{\epsilon_1}{\beta T}\left[\frac{4}{1-e^{-\beta T}}
\sinh^2\left(\frac{\beta T_l}{2}\right)+Q(\tau)\right], \\[4mm]
Q(\tau)=
\cases{
e^{\beta\tau}-e^{\beta T_l} & \text{if $0<\tau<T_l$,} \\
0 & \text{if $T_l<\tau<T-T_l$,} \\
e^{\beta(T-T_l)}-e^{\beta\tau} & \text{if $T-T_l<\tau<T$.} \\
}
\end{eqnarray}
\endnumparts

It is straightforward to check that $K_1$ in \eqref{eq:amItau} vanishes,
so the first term that may not be zero is $K_3$. Lengthy calculations
lead to
\begin{equation}
K_3=-\frac{32\epsilon_1^3}{\beta^4T^3}\,\frac{e^{\beta T}}{(e^{\beta T}-1)^2}
\sinh^2\left(\frac{\beta(T-2T_l)}{2}\right)\sinh^4\left(\frac{\beta T_l}{2}\right),
\label{eq:K3}
\end{equation}
that is to say
\begin{equation}
v=\frac{4\epsilon_1^3}{(\beta M)^3\beta T}
\sinh^2\left(\frac{\beta(T-2T_l)}{2}\right)\frac{
\sinh^4\left(\beta T_l/2\right)}{\sinh^2\left(\beta T/2\right)}
+o(\epsilon_1^3).
\label{eq:vsqwv}
\end{equation}
It is interesting to noticing that $K_3=0$ if $T_l=T/2$ because in that case
the  time-periodic piecewise constant force
 \eqref{sq-wv} is shift-symmetric. On the other hand, we can
determine the value of $T_l$ for which the ratchet effect is maximum by
differetiating \eqref{eq:vsqwv}. This leads to
\begin{equation}
\sinh\left(\frac{\beta(T-3T_l)}{2}\right)
\sinh\left(\frac{\beta(T-2T_l)}{2}\right)
\sinh^3\left(\frac{\beta T_l}{2}\right)=0.
\end{equation}
The only three solutions to this equation are $T_l=0$, $T_l=T/2$ and $T_l=T/3$.
The first two do not produce any ratchet current (with 
$T_{l}=0$ $f=0$ whereas for $T_{l}=T/2$ the force is shift-symmetric), 
therefore the last one
provides its maximum value.

As a final remark, expression \eqref{eq:vsqwv} has well defined overdamped
($M\to 0$, $\beta\to\infty$, with finite $\gamma=\beta M$) and undamped
($\beta\to 0$) limits. In fact, the undamped limit of \eqref{eq:vsqwv} yields 
\begin{equation}
v=\frac{\epsilon_1^3}{4(MT)^3}T_l^4(T-2T_l)^2+o(\epsilon_1^3),
\end{equation}
whereas the overdamped produces $v=o(\epsilon_1^3)$. Indeed, since
$f(t)^{2k+1}=\epsilon_1^{2k}f(t)$, in the overdamped case Eq.~\eref{eq-ov1}
immediately implies $v=0$. This is in marked contrast with the
overdamped deterministic
dynamic of a particle in a sinusoidal potential driven by a bi-harmonic 
force~\cite{cubero:2010}. In this case, the zero ratchet velocity
can be explained as a symmetry effect. Indeed, notice that $f(t)=-f(-t)$
when $f(t)$ is given by \eqref{sq-wv} (something that only happens for
the bi-harmonic force \eqref{eq-f} for specific choices of the phases),
and that the overdamped limit of Eq.~\eqref{equxa} remains invariant
under a simultaneously action of time-reversal and a sign change of
$x$ and $u$ (see \cite{quintero:2010} for further details).

\section{Discussion}
\label{sec:discussion}

We have studied the dynamics of a damped relativistic particle under two zero
average $T$-periodic forces which breaks the shift-symmetry $f(t+T/2)=-f(t)$. 
This nonlinear system can be explicitly solved through a transformation
that renders it linear. Therefore, the ratchet average velocity, $v$, is
exactly obtained for any arbitrary force $f(t)$.
This result allows us to show, first of all, that the ratchet velocity cannot
be obtained in general by using the method of moments (according to which
$v$ is obtained as a series of the odd moments of $f(t)$). And secondly,
that $v$ is a functional of $f(t)$, i.e. $v[f]$. Indeed, for any $T$-periodic 
force we have explicitly found the coefficients of the functional Taylor expansion
\eqref{eq-ov2}. In particular, this expansion shows that the method of
moments is only justified in the strict overdamped limit (see Eqs.~\eqref{eq-ov1}
and \eqref{eq-ov3}). Due to the symmetry $v[f]=-v[-f]$ only odd terms
contribute to the Taylor expansion. Besides, since the ratchet velocity
is translationally invariant, the kernel $c_{1}(t_{1})$ must be a
constant. So the first order term vanishes because the force
has zero average. Therefore, the first term in the expansion \eqref{eq-ov2} 
that is not necessarily zero is the third one, irrespective of the kind
of nonlinearity of the system. 

We have chosen to illustrate this functional representation 
the bi-harmonic force \eqref{eq-f} (with $p=2$ and $q=1$)
as well as a time-periodic piecewise constant force \eqref{sq-wv}.
We have obtained the leading term of the average velocity for both
these forces. They are given by equations~\eqref{vmedia-1}--\eqref{coeff}
and \eqref{eq:vsqwv}, respectively. It is worth emphasizing that the method
of moments always predicts a zero ratchet velocity when the system is
driven by a time-periodic piecewise constant. This is to be compared
with the result \eqref{eq:vsqwv} obtained here.
We have discussed the two limiting dynamics: undamped and overdamped.
In these two limits the system remains invariant if the driving force has the
symmetries $f(t)=f(-t)$ and $f(t)=-f(-t)$, respectively.
If the relativistic particle is driven by a bi-harmonic force
$v \sim \epsilon_{1}^{2} \epsilon_{2} \cos(2 \phi_{1}-\phi_{2})$ in the
overdamped limit, whereas $v \sim \epsilon_{1}^{2} \epsilon_{2}
\sin(2 \phi_{1}-\phi_{2})$ in the undamped limit. In the latter case
this means no ratchet current if we set $\phi_{1}=\phi_{2}=0$. The 
unexpected consequence of this is that introducing damping \emph{generates}
a ratchet current, whose intensity grows up to a maximum before it drops
to zero upon a further increase of the damping. This effect is a result
of a trade off between symmetry effects and friction and our prediction
is that should be observed in any system with damping and forced with
a time-reversible external force.
   
On its side, if an overdamped relativistic particle is driven by a
time-periodic piecewise constant force like \eqref{sq-wv}, the ratchet
velocity is always zero as a consequence of the symmetry $f(t)=-f(-t)$
exhibited by this force.

Summarizing, we hope to have illustrated the predictive power of the
Taylor functional expansion method introduced in \cite{quintero:2010}.
This working example also shows that this is the only reliable method
to analyze the ratchet current as a function of the parameters of the
external force. The most widely used alternative so far, the method
of moments, is shown to work only in the overdamped limit of the
dynamics of a relativistic particle driven by a periodic force. When
damping is finite and the forcing of the system is bi-harmonic, the
rathet current predicted by the method of moments still retains some
relevant features of the exact one \eqref{eq:vrelat}. However, it
dramatically fails if the system is driven by a piecewise constant
force, because it always predicts a zero ratchet current, in marked
constrast with the result \eqref{eq:vsqwv} predicted by the functional
Taylor expansion.

\ack

We acknowledge financial support through grants MTM2009-12740-C03-02 (R.A.N.),
FIS2008-02380/FIS (N.R.Q.), and MOSAICO (J.A.C.) (from Ministerio de Educaci\'on y
Ciencia, Spain), grants FQM262 (R.A.N.), FQM207 (N.R.Q.), and P09-FQM-4643
(N.R.Q., R.A.N.) (from Junta de Andaluc\'{\i}a, Spain), and project MODELICO-CM
(J.A.C.) (from Comunidad de Madrid, Spain). 


\section*{References}

\bibliographystyle{unsrt}
\bibliography{ratchets}

\end{document}